\documentstyle[aps,prb,preprint]{revtex}
\begin{document}
\title{Spin-Lattice Relaxation in Si Quantum Dots}
\author{B. A. Glavin}
\address{Institute of Semiconductor Physics, Ukrainian National
Academy of Sciences, Pr. Nauki 45, Kiev 03028, Ukraine}
\author{K. W. Kim}
\address{Department of Electrical and Computer Enginering, North Carolina
State University, Raleigh, NC 27695-7911, USA}

\maketitle

\begin{abstract}
We consider spin-lattice relaxation processes for electrons
trapped in lateral Si quantum dots in a $[001]$ inversion layer.
Such dots are characterized by strong confinement in the direction
perpendicular to the surface and much weaker confinement in the lateral
direction. The spin relaxation is assumed to be due to the modulation of
electron $g$-factor by the phonon-induced strain, as was shown previously for
the shallow donors. The results clearly indicate that the specific valley
structure
of the ground electron state in Si quantum dots causes strong anisotropy
for both the one-phonon and two-phonon spin relaxation rates. In addition, it
gives rise to a partial suppression of the two-phonon relaxation
in comparison to the spin relaxation of donor electrons.
\end{abstract}


\section{Introduction}

Recently, there is a growing interest in the physics of electron spin due to
the enormous potential of spin-based devices.  In these so-called "spintronic"
devices, information is encoded in the spin state of individual electrons.
Numerous concepts ranging from spin analogs of conventional electronic devices,
to quantum computers (which utilize the Zeeman doublet of a confined electron
as a qubit)~\cite{dotcomp} have been proposed.
Electron spin states relax by scattering with
imperfections or elementary excitations such as phonons.  Hence, the spin
relaxation time is a vital characteristic that determines the potential value of
a spin-based device.

In the present paper, we calculate the longitudinal spin-lattice relaxation
time ($T_1$) for electrons confined in Si quantum dots (QDs). Being a
relatively light semiconductor, Si is characterized by a weak
spin-orbit interaction, which basically determines the strength
of spin relaxation. In addition, only a small fraction of the isotopes in
natural Si possess a non-zero nuclear magnetic moment.  As a result,
the electron-nuclear
spin-flip process is expected to be slow. These two properties of Si,
along with the tempting possibility of integrating the "quantum" part
of a computer with the well-developed Si "classical" electronics,
make Si an attractive material for spin devices.

We concentrate on a particular design of QD based
on a $[001]$ inversion layer formed at the interface of Si and SiO$_2$
[see Fig.~1(a)].  The lateral confinement of electrons is assumed to be
due to the attractive potential applied to the gate electrode deposited
on top of the oxide layer. Alternative QD design based on Si/SiGe
heterostructures is also possible.  However, the SiGe-based design may face
additional complications for quantum computing applications. For example,
electron confinement in Si/SiGe heterostructures is relatively weak and
the penetration of electron wavefunction to the SiGe barriers is inevitable.
Since spin-orbit interaction in Ge is stronger than in Si,
these structures are expected to have higher spin relaxation rates.

In our calculations, we take advantage of the results obtained for spin
relaxation of shallow-donor electrons in
Si.\cite{feher,honig,hasegawa,roth1,roth2}
In Refs.~\onlinecite{hasegawa,roth1,roth2}, the spin-lattice relaxation
was assumed to be due to the modulation of electron $g$-factor by
the phonon-induced strain.
The parameters of the effective Hamiltonian were estimated from
microscopic models. Later, they were also determined by
measuring the $g$-factor under a static strain.\cite{feher}
Taking into account the structure of the electron states
in lateral QDs, we calculate the corresponding spin relaxation time
for both single-phonon and two-phonon processes. Simple
analytical expressions are obtained for the two
limiting cases, namely, for $T \ll E_Z$ and $T \gg E_Z$,
where $T$ is the temperature in energy units
and $E_Z$ is the Zeeman doublet splitting energy.
The former case is
more relevant to the conditions for quantum computation,
while the latter is close to the typical conditions of EPR
measurements. For the resonance frequency about $50~GHz$
and low enough temperature, the spin relaxation
time is found to be several minutes with the single-phonon
process providing the prevailing contribution to relaxation.
For elevated temperatures, the two-phonon process becomes significant and
the relaxation time can be substantially smaller. We also predict
strong anisotropy in the relaxation rate. In particular,
for the $[001]$ inversion layer, the spin  relaxation is suppressed if the
applied magnetic field
is parallel to the $[001]$ or $[110]$ directions.

The rest of the paper is organized as follows: In Section~II, we introduce the
Hamiltonian for spin-lattice relaxation and provide expressions for
the single-phonon and two-phonon relaxation rates. In Section~III,
the asymptotic dependences for different temperatures as
well as numerical results are presented. Finally, Section~IV is
devoted to the discussion of the obtained results along with other
potential mechanisms of spin-lattice relaxation.

\section{Effective Hamiltonian for Spin-Lattice Relaxation}

For crystals which possess inversion symmetry like Si, there is
no analog of conventional deformation potential for the spin-flip process,
which is often called Van Vleck cancellation.\cite{vleck}
This is a direct result of the requirement for the Hamiltonian to be
invariant under a generalized inversion transformation $C=JK$, where
$J$ and $K$ are  spatial and time inversion operators. In
Refs.\onlinecite{hasegawa,roth1,roth2}, the following
effective $C$-invariant Hamiltonian
describing the modulation of electron $g$-factor by strain
has been proposed:
\begin{equation}
\label{eq:n1}
H_g = A_{ijkl} u_{ij} B_k \sigma_l,
\end{equation}
where $A_{ijkl}$ are the coefficients, $u_{ij}$ is the strain tensor,
$B_k$ are the components of the magnetic field, and $\sigma_l$ are the Pauli
matrices;
here and below we assume summation over the repeated indices.
$H_g$ is written in the basis of the Bloch functions corresponding to the
bottom of the conduction band and can be used for calculations
of the spin transitions for the electron states described within the
effective mass approximation.
Nonzero coefficients $A$ are determined by the symmetry of
the crystal using the method of invariants.\cite{bir-pikus}
In particular, for the $\Delta$ point of the Brillouin band in a diamond-like
crystal that corresponds to the conduction band of
Si, there are eight invariants and the Hamiltonian
for a $[001]$ valley can be written as:\cite{sheka}
\begin{eqnarray}
\label{eq:n2}
H_g^{[001]} = \frac{1}{2} \mu_B \left( A_1 \sigma_z B_z (u_{xx} +u_{yy})+
A_2 \sigma_z B_z u_{zz}+ \right.\\
A_3 \sigma_z (B_x u_{xz} + B_y u_{yz}) +
A_4 (\sigma_x B_x +\sigma_y B_y) u_{zz} + \nonumber \\
A_5 (\sigma_x B_x +\sigma_y B_y) (u_{xx}+ u_{yy})+
A_6 (\sigma_x B_y +\sigma_y B_x) u_{xy} + \nonumber \\
\left. A_7 (\sigma_x u_{xz} +\sigma_y u_{yz} ) B_z +
A_8 (\sigma_x B_x - \sigma_y B_y ) (u_{xx} - u_{yy})\right). \nonumber
\end{eqnarray}
Here, we introduced the factor $\mu_B/2$ for convenience,
$\mu_B$ being the Bohr magneton.
The absolute values of the coefficients $A$ can be determined
in principle by using a many-band effective-mass
expansion of the electron wavefunction in
a uniform magnetic field, similar to that used in Ref.\onlinecite{roth1}.
In practice, this cannot be accomplished
since the required momentum matrix elements are unknown.
However, some qualitative considerations are possible.
Since the terms of the expression for $g$-factor contain the
energy gaps between the coupled bands in the denominator,
coupling of the closest bands is expected to be the strongest.
For Si, there is a $\Delta'_2$ band which is close to the $\Delta_1$
conduction band.  These bands merge at the $X$ point in the momentum space
but are not coupled by either a spin-orbit interaction or
momentum operators. Therefore, $\Delta_1 \leftrightarrow \Delta'_2$
coupling is not manifested in the effective mass or $g$-factor of
unstrained Si. However,
these bands are coupled by the deformation potential. From the character
tables of the $\Delta$ point and the corresponding invariants, it is easy to
conclude that this coupling is realized by the deformation potential term
proportional to $u_{xy}$. Since $H_g$ ccontains
 only one invariant proportional
to $u_{xy}$, it can be concluded that the major contribution
to the Hamiltonian is
\begin{equation}
\label{eq:n3}
H_g^{[001]} = \frac{1}{2} A \mu_B (\sigma_x B_y +\sigma_y B_x) u_{xy},
\end{equation}
where we drop the index of the coefficient $A_6$. This argument was initially
used by Roth.\cite{roth2}

To proceed with the calculation of the longitudinal relaxation time
$T_1$, we need to describe explicitly the system under consideration
as well as the characteristic energy and length scales.
As mentioned in the introduction, we consider Si lateral QDs formed
at the Si/SiO$_2$ interface, where the
lateral confinement is due to the gate electrodes.
For such a system, the lateral dimension of the QD $a_{lat}$ is typically on
the order of one hundred $nm$, exceeding considerably
the inversion layer thickness $a_{2D}$.
The energy structure of the electron levels
is determined by the following parameters: (a)~quantization energy in the
2D inversion channel $E_{2D} \sim  \hbar^2 / (2 m_{2D} a_{2D}^2)$;
(b)~lateral quantization energy $E_{lat} \sim \hbar^2 / (2 m_{lat}
a_{lat}^2)$; (c)~intervalley splitting energy between 
$[001]$ and $[00\bar{1}]$ states, $\Delta$; and (d)~Zeeman energy
$E_Z =  \mu_B g B$, where $g$
is the effective $g$-factor of the confined electrons depending, in general,
on the direction of the magnetic field.
Here $m_{2D}$ and $m_{lat}$ are the electron effective masses
in the direction normal and parallel to the inversion layer,
respectively. For the $[001]$ inversion layer, they just correspond
to the longitudinal and transverse effective masses of Si.
In the following, we assume that the conditions
$E_Z \ll \Delta,\, E_{lat}$ and $ E_{lat} \ll E_{2D}$ are satisfied.
In fact, this is a necessary requirement for spin qubit operation.
Using modern technology, $E_{lat}$ can be made about a $meV$ or even higher.
In contrast, $\Delta$ can be controlled in a much lesser degree, and
is roughly proportional to the confining electric field in an inversion
layer.  Experimentally, $\Delta$ was measured to be
up to $1~meV$ in strong fields (see Ref.\onlinecite{ando}).
In Fig.\ 1(b), we show schematically the energy levels and the electron
transitions under consideration. The numbers "0" and "1" mark the levels of
lateral electron confinement.
The signs $+$ and $-$ denote the valley-split
electron states, and ``up'' and ``down'' are the spin states.
We assume that only the lowest Zeeman doublet can be populated, which
means that the temperature $T$ is much less than $\Delta$ and
$E_{lat}$. The longitudinal
relaxation time $T_1$ is determined as $T_1^{-1} = W_{up-down} +
W_{down-up}$, where $W$ are the probabilities of spin-up to spin-down
and spin-down to spin-up transitions. The solid arrows indicate
the single-phonon transitions, while the dashed arrows correspond to
the two-phonon process. Later in this section we comment on the possible
two-phonon processes shown in the figure.

In Refs.\onlinecite{hasegawa} and \onlinecite{roth1}, an alternative
Hamiltonian to Eq.~(\ref{eq:n3}) was considered.  It originates from the
coupling of the donor singlet and doublet states by an applied
magnetic field. It arises
due to the different valley structure of these states and the anisotropy of
the $g$-factor in the individual valleys. Although this Hamiltonian
does not involve $\Delta \leftrightarrow \Delta'_2$ coupling,
its contribution is high because of a very small gap between the
singlet and doublet states. However, there is no such mechanism in the
$[001]$ lateral QDs, which is probably the main difference with
the case of spin relaxation for donor electrons.
The reason for that can be easily seen.  For the $[001]$ inversion layer,
the ground electron state is a combination of $[001]$ and
$[00\bar{1}]$ states:
\begin{equation}
\label{eq:n4}
\Psi_{\pm} = \chi  \left(
F_{001}^\pm \psi_{001} + F_{00\bar{1}}^\pm \psi_{00\bar{1}}\right),
\end{equation}
where we dropped the spin index of wavefunctions.
Here $\chi$ is the envelope wavefunction, $\psi_{001}$ and
$\psi_{00\bar{1}}$ are the Bloch functions corresponding to
the $[001]$ and $[00\bar{1}]$ valleys, respectively, and $F$ are
the coefficients which determine the valley splitting.
Particular expressions for $F$ can be found using a microscopic
model, for example, that of Sham and Nakayama.\cite{sham}
In our case we do not need explicit expressions for
$C$. It is enough to use the fact that Zeeman Hamiltonian has
identical forms for the $[001]$ and $[00\bar{1}]$ valleys and
its intervalley matrix elements are zero.
Therefore, the matrix element between $\Psi_+$ and $\Psi_-$, which is
proportional to the overlap between them, is zero because these states
are orthogonal.

The Hamiltonian of Eq.~(\ref{eq:n3}) is written in the representation
where the basis electron wavefunctions correspond to the definite
spin projections on the $z$ axis. For calculation
of $T_1$, it must be rewritten by using a representation with
the definite spin projection on the direction of the magnetic
field.\cite{note1}
This can be done following the standard procedure of transformation for
Pauli matrixes under rotation.\cite{landau}
Finally, we obtain the expression for the one-phonon
relaxation rate, $1/T_1^{(1)}$ as:
\begin{equation}
\label{eq:n5}
\frac{1}{T_1^{(1)}} = \frac{\pi^3 A^2}{4} \frac{\hbar f^5}{g^2 \rho}
(1+2N_T) \sin^2 \theta (\cos^2 2\phi +\cos^2\theta \sin^2 2\phi)
\sum_i \int d\Omega_q^{(i)} \frac{(e_x^{(i)}n_y^{(i)} +
e_y^{(i)} n_x^{(i)} )^2}{s_i^5}.
\end{equation}
In this equation, $g$ is the slightly anisotropic $g$ factor of confined
electrons, $f=g\mu_B B/(2\pi \hbar)$ is the resonance frequency,
$\theta$ and $\phi$ are the spherical angles of the magnetic field,
$\rho$ is the material density, $N_T$ is the Planck phonon population for
the energies equal to $E_Z$, $\Omega_q$ is a solid angle in the phonon
wavevector space {\boldmath $q$},
{\boldmath $e$} and {\boldmath $n$} are the phonon
polarization vector and the unit vector parallel to {\boldmath $q$},
respectively, $s$ is the sound velocity, and the summation is over
the acoustic phonon branches. Equation~(\ref{eq:n5}) assumes
that the phonon wavelength corresponding to the energy $E_Z$ is much
greater than the lateral dimensions of QD.
In this case, the
form factor of the electron-phonon interaction is equal to unity and
$T_1^{(1)}$ does not depend on the particular shape of the lateral
confining potential.
To check the validity of this approach, we performed
calculations of $T_1^{(1)}$ assuming parabolic lateral confinement
and found that for the lateral level separation of $1~meV$, the
obtained correction is less than $10~\%$ even for $f=50~GHz$.
An additional assumption of the bulk-like phonon spectrum is made for
simplicity.
This probably introduces a greater error since the lateral QDs are normally
situated close to the surface. As shown in
Refs.\onlinecite{levinson,sirenko,pipa}, the phonon modes in this case
are essentially rebuilt, due to the interference of the incident and
reflected phonons as well as the origination of Rayleigh waves, which strongly
modify electron-phonon coupling.

Equation~(\ref{eq:n5}) predicts strong anisotropy of the relaxation rate.
If the magnetic field is parallel to the $[001]$ or $[110]$ directions,
$T_1^{(1)}$ goes to infinity. This is because we used a single-parameter
Hamiltonian of Eq.~(\ref{eq:n3}). When rewritten for the basis with
the definite spin projection on the direction of the magnetic field,
it vanishes for these particular orientations. Of course, in experiments
the relaxation for these cases is not expected to be suppressed completely,
since the remaining terms of Eq.~(\ref{eq:n2}) have non-zero contributions to
the relaxation rate. However, a significant decrease is expected.

Assuming an isotropic acoustic phonon spectrum, we obtain
\begin{equation}
\label{eq:n6}
\frac{1}{T_1^{(1)}} = \frac{2\pi^4 A^2}{5} \frac{\hbar f^5}{g^2 \rho s_t^5}
(1+2N_T) \sin^2 \theta (\cos^2 2\phi +\cos^2\theta \sin^2 2\phi).
\end{equation}
Here we take into account only TA phonons, which provide
the major contribution to the relaxation rate.

Let us now turn to the calculation of the relaxation time due
to the two-phonon transitions $T_1^{(2)}$. This is a second-order
transition where the electron is virtually scattered first to an
intermediate state and then to the final state.
One of the virtual transitions is accompanied by
spin flip, while the other occurs with spin conservation.
The probability of a two-phonon transition can be found using
the second-order perturbation theory (see, for example,
Ref.\onlinecite{gantmakher}). For the spin relaxation of donor electrons,
the intermediate electron state is represented by the excited doublet.
In contrast, for the case of $[001]$ lateral QD,
the valley-split state cannot serve as an intermediate state. This
is because the valley-split states are not coupled by the
Hamiltonian of Eq.~(\ref{eq:n2}). This can be shown using a argument
similar to that applied for the proof of the absence of $g$-factor
modulation due
to coupling of the valley-split states. Other possible transitions
are through the excited states of the lateral confinement. Since the
intervalley splitting is controlled by the electron
confinement in the $z$ direction rather than in the lateral direction,
these transitions are actually possible between the states having the
same valley structure [see Fig.~1(b)]. If the phonon wavevector is very small,
then the probability of such a transition goes to zero because the overlap
functions of "0" and "1" states are orthogonal. To determine the
probability of two-phonon transition for a finite
phonon wavevector, we need to know the overlap
function $\chi$ explicitly. For calculations, we assume parabolic lateral
confinement. In this case, the "lateral" electron state is
determined by the two quantum numbers, $l_x$ and $l_y$. The "0"
state corresponds to $l_x=0$, $l_y =0$, and there are two
degenerate "1" states with $l_x=0$, $l_y=1$, and $l_x=1$,
$l_y=0$. Using the expressions for the wavefunctions of harmonic
oscillator,\cite{landau} it is easy to obtain the necessary form-factor
$J$:
\begin{equation}
\label{eq:n7}
J \equiv  \int dx\, dy\, dz\, \chi_{00} \chi_{10} \exp \left(i (q_x x
+ q_y y +q_z z)\right)  = \frac{1}{\sqrt{2}} \frac{q_x}{k}\exp
\left(-\frac{q_x^2 +q_y^2}{4k^2}\right),
\end{equation}
where the subscript of $\chi$ represent the $l_x$ and $l_y$ quantum numbers
and $k$ is expressed through the energy gap $\delta$ between "0" and
"1" states and the lateral effective mass:
$k=\sqrt{m_{lat}\delta}/\hbar$. Here we take into account
that the thickness of
the inversion layer is much less than the typical phonon wavelength.
In the following, we also assume that the typical phonon wavevector is
less than $k$, and drop the exponent in the expression for $J$.
Note, that $J$ can be modified due to the diamagnetic
influence of the magnetic field as well. In particular,
it can lift the degeneracy of
"1" states. In our calculations, we do not consider this effect.

We can distinguish three contributions to the two-phonon relaxation rate:
(a)~due to emission of two phonons, (b)~due to absorption of two phonons,
and (c)~due to phonon scattering. For each of them the rate can be
expressed in a uniform manner:
\begin{equation}
\label{eq:n8}
\frac{1}{T_1^{(2)}} = 34 \pi^{10} \left(\frac{16}{105}\right)^2
\frac{A^2 E_2^2\hbar^6 f^{13}}{\delta^4 g^2 \rho^2 s_t^{14} m_{lat}^2}
\sin^2 \theta (\cos^2 2\phi +\cos^2\theta \sin^2 2\phi) D_i.
\end{equation}
Here only the transverse phonons are
taken into account, which provide the
major contribution to the relaxation rate. We also assume that
typical phonon energies are considerably less than $\delta$.
The spin-conserving virtual transition is treated within the
deformation model, which for the $[001]$ valley provides
the interaction  $E= E_1 u_{ii} +
E_2 u_{zz}$, where $E_1$ and $E_2$ are the deformation
potential constants. The coefficients $D$ depend on temperature and they
are different for each of the three processes mentioned:
\begin{eqnarray}
\label{eq:n9}
D_{em}=\int_0^1 dx\, x^5 (1-x)^5 \left(
1 + \frac{1}{\exp (x/t) -1}\right) \left(
1 + \frac{1}{\exp ((1-x)/t) -1}\right),\\
D_{ab}=\int_0^1 dx\, x^5 (1-x)^5 \frac{1}{\exp (x/t) -1}
\, \frac{1}{\exp ((1-x)/t) -1}, \nonumber \\
D_{scat} = \frac{20}{17} \left(\int_0^\infty dx\, x^5 (1+x)^5
\frac{1}{\exp (x/t)-1} \left( 1+ \frac{1}{\exp ((1+x)/t)-1}\right) +
\nonumber \right.\\
\left. \int_0^\infty dx\, x^5 (x+1)^5
\frac{1}{\exp ((x+1)/t)-1} \left( 1+ \frac{1}{\exp (x/t)-1}\right)\right),
\nonumber
\end{eqnarray}
where $t=T/E_Z$. One can see, that the two-phonon relaxation rate is
characterized by the same anisotropy as in the one-phonon relaxation rate.
In the following section, we analyze the asymptotic dependences
of the relaxation rates for different temperature regimes and perform
numerical calculations.

\section{Relaxation Rates for Low and High Temperatures: Numerical Results}

For the limiting cases of $T \ll E_Z$ and $T \gg E_Z$, the relaxation
rates obey simple power laws as a function of resonance frequency
and temperature. In particular,
$T_1^{(1)} \sim f^{-5}$ in the former case and $T_1^{(1)} \sim
f^{-4}T^{-1}$ in the latter case. This is similar to the case of
donor spin relaxation.\cite{hasegawa,roth1,roth2}
The two-phonon relaxation time for $T \ll E_Z$ is proportional to
$f^{-13}$ with the main contribution from the two-phonon emission.
For $T \gg E_Z$, the relaxation is mainly due to the phonon scattering and
$T_1^{(2)} \sim f^{-2} T^{-11}$. For donor electrons,
Roth obtained a different power law.\cite{roth1} This is because
for donor electrons the form-factor of singlet-doublet transition in
the lowest approximation is equal to unity,
in contrast to Eq.~(\ref{eq:n7}) for the lateral QD where
the form-factor is suppressed for long-wavelength phonons.

In Figs.~2 and 3, we show the results of numerical calculations for
one-phonon and two-phonon rates. We assume a magnetic field parallel to
the $[100]$ direction, $\rho =2329~kg/m^3$,
the transverse sound velocity $s_t=5420~m/s$, $E_2 =10~eV$,
$m_{lat} = 0.19~m_0$, $\delta =2~meV$, and $g=2$.
The coefficient $A$ can be determined by the
measurement of the donor $g$-factor in strained Si.
With this method, $A=1.32$ was obtained in Ref.\onlinecite{feher}.
The dependence of the spin relaxation rate on the resonance
frequency is presented in Fig.~2 for several temperatures. One can see that
for $T_1^{(1)}$ the transition between the described power dependences
at the different temperature regimes is quite fast.
This is not the case for $T_1^{(2)}$, which is due to
the different temperature dependence of phonon absorption,
emission, and scattering rates. 

In Fig.~3, we plot the relaxation rates as a function of $f$
under the condition $T/E_Z =const$. This is relevant to the
case of QD-based qubit, where this ratio must be kept small
to ensure initial state preparation of the qubits.
We see that under this condition the major contribution is provided
by one-phonon scattering.

Of course, the predicted for small resonance frequencies 
 huge relaxation times hardly can be measured in experiments.  
This is similar to the case of donor spin relaxation,\cite{feher}
where actual experiments were undertaken for $f$ about  tens of $GHz$.

\section{Discussion}

Let us first summarize the major distinction of the spin-relaxation
process for lateral QDs and shallow donors. First, we predict that
the relaxation for lateral QDs is more anisotropic than that for
donor electrons. In particular, for $[001]$ inversion layer
 the relaxation rate is suppressed
for a magnetic field parallel to the $[001]$ or $[110]$ direction.
Second, the two-phonon relaxation in lateral QDs is, in general, weaker
than for donor electrons and is characterized by different power dependences
on the resonance frequency and temperature.
Both of these features arise due to the different valley structure of
the electron states in lateral QDs in comparison to that of donor electrons.

We have to stress that the first conclusion relies on the model
used for calculations, which must be checked by experiments.
This is because the values of the coefficients $A_n$ in Eq.~(\ref{eq:n2})
are determined not only by the energy gaps between the bands, but also by
a number of interband matrix elements, which are unknown.
Strictly speaking, the experiments with donors can not be
considered as a rigorous proof of single-valley Hamiltonian
of Eq.~(\ref{eq:n3}). In fact, for the donor electrons, eight invariants
of Eq.~(\ref{eq:n2}) are transformed to three invariants after
summation over the valleys. The term measured in Ref.\onlinecite{feher},
$H \sim u_{xy} (\sigma_x B_y +\sigma_y B_x)) +cp$ where $cp$ stands for
cyclic permutations, is obtained from several terms of Eq.~(\ref{eq:n2}),
not only from that of Eq.~(\ref{eq:n3}).

There is another mechanism that can modify the spin relaxation in
the lateral QDs.
This can take place if some quantum levels originated from
the longitudinal $[001]$, $[00\bar{1}]$ valleys and
the transverse $[100]$, $[\bar{1}00]$, $[010]$, $[0\bar{1}0]$
valleys come close. In this case, the intervalley coupling can mix these
two groups. Mixing of these states and the ground state of
the QD by both steady-state and strain-induced contributions of Zeeman
Hamiltonian can be possible. As a result, both one-phonon
and two-phonon relaxation
will be modified. According to self-consistent calculations, \cite{ando}
such a situation is possible for
particular parameters of Si inversion layer.

Finally, it is necessary to take into account alternative forms of the
Hamiltonian which also cause spin relaxation. For example, the following
Hamiltonian
is possible:
\begin{equation}
\label{eq:n10}
H_{sp} = D_{ijkl} \frac{\partial u_{ij}}{\partial r_k} \sigma_l.
\end{equation}
As with $H_g$, this Hamiltonian is invariant
under inversion $C$. Physically, it describes the splitting of the spin states
by nonuniform deformation. A similar relaxation process proportional to
the third power of the phonon wavevector, was discussed previously in
Ref.\onlinecite{yafet}. We believe that the disagreement between the
theoretical
calculations of $T_1$ based on $H_g$ and experimental
measurements \cite{feher} can be partially explained by the contribution of
$H_{sp}$. Indeed, it is easy to check that $H_{sp}$ provides
the same dependence of the one-phonon rate on frequency and temperature
as $H_g$. This is because the energy conservation makes the magnetic field and
the phonon wavevector proportional to each other. In Ref.\onlinecite{feher},
the measured one-phonon rate was about two times higher than the calculated one.
The two-phonon relaxation was found to obey the predicted temperature
dependence, but in contradiction to $H_g$,
demonstrated no dependence on the magnetic field. Assuming that the
contribution of $H_{sp}$ to one-phonon relaxation is about that of $H_g$, the
contribution of $H_{sp}$ to two-phonon rate in this case can be much greater
than that of $H_{g}$ since the typical phonon energy for two-phonon process
is much greater than that of Zeeman splitting. Hence, this assumption can
explain both the quantitative disagreement obtained for one-phonon relaxation
and the qualitative one obtained
for two-phonon relaxation. Of course, this idea must be checked
experimentally. In particular, important information can be obtained
from the measurement of possible spin splitting in nonuniformly strained
Si. Consideration of the symmetry properties and
the absolute value of such splitting will be provided elsewhere.

Finally, we would like to stress that the electron spin relaxation in QDs
is of the same order of magnitude as that obtained for donors. It is much
longer than the $T_1$ in III-V compounds, which proves a good perspective of
Si for quantum information processing.

\section*{Acknowledgements}
The work performed at North Carolina State University was supported by the
Office of Naval Research and the Defense Advanced Research Projects Agency.

\begin{figure}
\caption{
(a) Schematics of the lateral QD. Confinement in the $[001]$ direction is
achieved as in the conventional inversion layers, while the lateral confinement
is provided by the electrodes (black boxes). (b) Schematic illustration of
energy levels in lateral QDs. (0) and (1) mark the different states of
lateral confinement. Valley-split $+$ and $-$ states are further
split by the magnetic field into spin-up and spin-down states.
The solid (dashed) arrows show electron transitions under one-phonon
(two-phonon) relaxation.
}
\end{figure}

\begin{figure}
\caption{
Dependence of one-phonon (solid lines) and two-phonon
(dashed lines) spin relaxation rates on
resonance frequency for temperatures of  $0.5~K$, $1~K$, and $2~K$.
}
\end{figure}

\begin{figure}
\caption{
Dependence of the spin relaxation rates on resonance frequency
for the case where the ratio of temperature and Zeeman splitting
$t$ is kept constant.
The one-phonon rates for the considered values of this ratio
cannot be resolved on this scale and is presented by one curve.
}
\end{figure}


\begin{thebibliography}{99}
\bibitem{note1}
It is worth noting
that such wavefunctions are not the eigenfunctions of the Hamiltonian
when  spin-lattice coupling is not considered. This is a result of the
anisotropy in the electron $g$-factor. For Si, this anisotropy is very weak
and is not taken into account.
\end{thebibliography}
\end{document}